\documentstyle[amssymb,aps]{revtex}

\begin{document}
\title{Interacting spin 0 fields with torsion via Duffin-Kemmer-Petiau theory}
\author{J. T. Lunardi\thanks{%
On leave from Departamento de Matem\'{a}tica e Estat\'{\i}stica. Setor de
Ci\^{e}ncias Exatas e Naturais. Universidade Estadual de Ponta Grossa. Ponta
Grossa, PR - Brazil.}, B. M. Pimentel and R. G. Teixeira}
\address{Instituto de F\'{\i}sica Te\'{o}rica,\\
Universidade Estadual Paulista.\\
Rua Pamplona, 145\\
01405-900-- S\~ao Paulo, SP\\
Brazil.}
\date{\today}
\maketitle

\begin{abstract}
Here we study the behaviour of spin 0 sector of the DKP field in spaces with
torsion. First we show that in a Riemann-Cartan manifold the DKP field
presents an interaction with torsion when minimal coupling is performed,
contrary to the behaviour of the KG field, a result that breaks the usual
equivalence between the DKP and the KG fields.

Next we analyse the case of Teleparallel Equivalent of General Relativity
(Weitzenb\"{o}ck manifold), showing that in this case there is a perfect
agreement between KG and DKP fields. The origins of both results are also
discussed.
\end{abstract}


\section{Introduction}

The Duffin-Kemmer-Petiau (DKP) equation is a first order relativistic wave
equation for spin 0 and 1 bosons \cite{Duffin,Kemmer,Petiau} given by 
\begin{equation}
\left( i\beta ^{a}\partial _{a}-m\right) \psi =0,  \label{eq1}
\end{equation}
where the matrices $\beta ^{a}$ obey the algebraic relations 
\begin{equation}
\beta ^{a}\beta ^{b}\beta ^{c}+\beta ^{c}\beta ^{b}\beta ^{a}=\beta ^{a}\eta
^{bc}+\beta ^{c}\eta ^{ba};  \label{eq2}
\end{equation}
where $a,b=0,1,2,3$, with $\eta ^{ab}$ being the metric tensor of Minkowski
space-time with signature $\left( +---\right) $. The latin alphabet will be
used all through this paper to indicate Minkowski indexes, while
Riemann-Cartan indexes will be indicated by greek letters.

The DKP equation is very similar to Dirac's equation but the algebraic
properties of $\beta ^{a}$ matrices, which have no inverses, make it more
difficult to deal with.

This equation can also be obtained from the Lagrangian density 
\begin{equation}
{\cal L}=\frac{i}{2}\overline{\psi }\beta ^{a}\stackrel{\leftrightarrow }{%
\partial }_{a}\psi -m\overline{\psi }\psi ,  \label{eq6}
\end{equation}
where $\overline{\psi }$ is defined\ as 
\begin{equation}
\overline{\psi }=\psi ^{\dagger }\eta ^{0},  \label{eq6b}
\end{equation}
and $\eta ^{0}$ is a matrix defined as 
\begin{equation}
\eta ^{a}=2\left( \beta ^{a}\right) ^{2}-\eta ^{aa}.  \label{eq3b}
\end{equation}
An equation for $\overline{\psi }$ can also be obtained from variational
principles from the above Lagrangian density or by hermitian conjugation if
we choose $\beta ^{0}$ to be hermitian and $\beta ^{i}$ $\left(
i=1,2,3\right) $\ anti-hermitian. If one uses two sets of Dirac matrices $%
\gamma ^{a}$ and $\gamma ^{\prime a}$ acting on different indexes of a 16
component $\psi $ wave function it can be verified that the matrices 
\begin{equation}
\beta ^{a}=\frac{1}{2}\left( \gamma ^{a}I^{\prime }+I\gamma ^{\prime
a}\right)   \label{eq6c}
\end{equation}
satisfy the algebraic relation (\ref{eq2}), but these matrices form a
reducible representation since it can be shown \cite{Kemmer,Umezawa} that
this algebra has 3 inequivalent irreducible representations: a trivial 1
degree $\left( \beta ^{a}=0\right) $ without physical significance; a 5
degree one, corresponding to a 5 component $\psi $ that describes a spin 0
boson; and a 10 degree one, corresponding to a 10 component $\psi $
describing a spin 1 boson.

Moreover, for any representation, one can define a set of \ ``projectors''
which select the scalar and vector sectors of the DKP field \cite{Umezawa}.
A detailed analysis of the DKP equation and of the properties of $\beta $
and $\eta $ matrices can be found in the original works \cite{Duffin,Kemmer}%
, while a similar analysis, complemented by the definitions of the
projectors for the scalar and vector sectors of the theory and their
properties, can be found in reference \cite{Umezawa}. A rather complete
introduction to DKP theory, that covers all properties necessary to the
reading of this work and uses the same metric signature used here, can be
found in references \cite{Lunardi 1} or \cite{Lunardi 2}. For a historical
development of this theory, among others, until the 70's reference \cite
{Krajcik} is suggested, while the application of DKP field to scalar QED\
can be found in \cite{Berestetskii}.

More recently there arose some new interest in DKP theory, specifically it
has been applied to QCD (large and short distances)\ by Gribov \cite{Gribov}%
, to covariant Hamiltonian dynamics \cite{Kanatchikov}; have been studied in
curved space-time \cite{Lunardi 1,Red'kov}, in Causal Approach \cite{Lunardi
3} and with five-dimensional Galilean covariance \cite{Esdras}. There also
have been given detailed proofs of the equivalence between DKP and KG fields
in various situations \cite{Fainberg 1,Fainberg 2,Fainberg 3} and some
points regarding DKP interaction with electromagnetic field have been
clarified \cite{Lunardi 2,Nowakowski}. In this work we will analyse the
interaction of the spin 0 sector of the DKP field with torsion in the cases
of Einstein-Cartan (EC) theory, which uses a Riemann-Cartan (RC) manifold to
describe gravitational interaction, and of the Teleparallel Equivalent of
General Relativity (TEGR).

\section{DKP field in Riemann-Cartan space-time}

To interact the DKP field with Riemann-Cartan space-time ${\cal U}^{4}$ we
introduce a tetrad set and follow the usual procedure used in the case of
Dirac's spinor. Details of this procedure can be found in literature and we
suggest specifically the reference \cite{Sabbata}, which notations and
definitions we follow. Bellow we just mention some basic identities and
results, for the sake of completeness. Moreover, the case of DKP field in
Riemannian space-times (i.e. torsionless case) was analysed by the authors
in reference \cite{Lunardi 1} and most of the results there showed are still
valid in the presence of torsion, such that they will be used here whenever
it is possible.

As usual the tetrad vector fields $e_{\mu }{}^{a}\left( x\right) $ satisfy,
at each point $x$ of ${\cal U}^{4}$, the relations 
\begin{equation}
\eta ^{ab}=e_{\mu }{}^{a}\left( x\right) e_{\nu }{}^{b}\left( x\right)
g^{\mu \nu }\left( x\right) ,\quad \eta _{ab}=e^{\mu }{}_{a}\left( x\right)
e^{\nu }{}_{b}\left( x\right) g_{\mu \nu }\left( x\right) ,  \label{eq21}
\end{equation}
\begin{equation}
g_{\mu \nu }\left( x\right) =e_{\mu }{}^{a}\left( x\right) e_{\nu
}{}^{b}\left( x\right) \eta _{ab},\quad g^{\mu \nu }\left( x\right) =e^{\mu
}{}_{a}\left( x\right) e^{\nu }{}_{b}\left( x\right) \eta ^{ab},
\label{eq22}
\end{equation}
the Latin indexes being raised and lowered by the Minkowski metric $\eta
^{ab}$ and the Greek ones by the metric $g^{\mu \nu }$ of the ${\cal U}^{4}$
manifold. The components $B^{ab}$ in the tangent Minkowski space-time ${\cal %
M}^{4}$ of a tensor $B^{\mu \nu }$ defined on ${\cal U}^{4}$ are given by 
\begin{equation}
B^{ab}=e_{\mu }{}^{a}e_{\nu }{}^{b}B^{\mu \nu },\ B_{ab}=e^{\mu
}{}_{a}e^{\nu }{}_{b}B_{\mu \nu }.  \label{eq26}
\end{equation}
The Lorentz covariant derivative $D_{\mu }$ is defined as 
\begin{equation}
D_{\mu }B^{a}=\partial _{\mu }B^{a}+\omega _{\mu }{}^{a}{}_{b}\left(
x\right) B^{b},  \label{eq28}
\end{equation}
\begin{equation}
D_{\mu }B_{a}=\partial _{\mu }B_{a}-\omega _{\mu }{}^{b}{}_{a}B_{b},
\label{eq30}
\end{equation}
where $\omega _{\mu }{}^{ab}$ is the connection on the tangent Minkowski
space-time, i.e. the {\it spin connection}.

The {\bf total} covariant derivative $\nabla _{\mu }$ of a quantity $B_{\nu
}{}^{a}$ will then be 
\begin{equation}
\nabla _{\mu }B_{\nu }{}^{a}=D_{\mu }B_{\nu }{}^{a}-\Gamma _{\mu \nu
}{}^{\alpha }B_{\alpha }{}^{a},  \label{eq31}
\end{equation}
or 
\begin{equation}
\nabla _{\mu }B^{\nu a}=D_{\mu }B^{\nu a}+\Gamma _{\mu \alpha }{}^{\nu
}B^{\alpha a},  \label{eq32}
\end{equation}
where $\Gamma _{\mu \nu }{}^{\alpha }$ is the {\it nonsymmetric} connection
on the ${\cal U}^{4}$ manifold given as 
\begin{equation}
\Gamma _{\mu \nu }{}^{\alpha }=\stackrel{r}{\Gamma }_{\mu \nu }{}^{\alpha
}-K_{\mu \nu }{}^{\alpha }=%
{\alpha  \atopwithdelims\{\} \mu \nu }%
-K_{\mu \nu }{}^{\alpha },  \label{eq32b}
\end{equation}
where $\stackrel{r}{\Gamma }_{\mu \nu }{}^{\alpha }=%
{\alpha  \atopwithdelims\{\} \mu \nu }%
$ is the connection of General Relativity given by the Christoffel symbols 
\begin{equation}
{\alpha  \atopwithdelims\{\} \mu \nu }%
=\frac{1}{2}g^{\alpha \beta }\left( \partial _{\mu }g_{\beta \nu }+\partial
_{\nu }g_{\beta \mu }-\partial _{\beta }g_{\mu \nu }\right) ,  \label{eq32c}
\end{equation}
and $K_{\mu \nu \alpha }{}$ is the contorsion tensor, antisymmetric in the
two last indexes, defined in terms of the torsion tensor\footnote{%
We use square brackets to designate antisymmetrization, i.e. $A_{\left[ \mu
\nu \right] }=\frac{1}{2}\left( A_{\mu \nu }-A_{\nu \mu }\right) $, and
round brackets to designate symmetrization, i.e. $A_{\left( \mu \nu \right)
}=\frac{1}{2}\left( A_{\mu \nu }+A_{\nu \mu }\right) $.} $Q_{\alpha \beta
}{}^{\mu }=\Gamma _{\left[ \alpha \beta \right] }{}^{\mu }$ as 
\begin{equation}
K_{\mu \alpha \beta }=-Q_{\mu \alpha \beta }-Q_{\beta \mu \alpha }+Q_{\alpha
\beta \mu }.  \label{eq32d}
\end{equation}

The derivatives above reduce to the Lorentz derivative $D_{\mu }$ given in
equation (\ref{eq28}) if the quantity $B$ has no ${\cal U}^{4}$ indexes and
to the usual Riemann-Cartan covariant derivative, i.e. 
\begin{equation}
\nabla _{\mu }B^{\nu }=\partial _{\mu }B^{\nu }+\Gamma _{\mu \alpha }{}^{\nu
}B^{\alpha },  \label{eq33}
\end{equation}
if $B$ has no ${\cal M}^{4}$ indexes. From the requirement that $\nabla
_{\mu }e_{\nu }{}^{a}=0$ we get a relation between the connections $\omega
_{\mu }{}^{ab}$ and $\Gamma _{\mu \nu }{}^{\alpha }$, i.e. 
\begin{equation}
\omega _{\mu }{}^{ab}=e_{\alpha }{}^{a}e^{\nu b}\Gamma _{\mu \nu }{}^{\alpha
}-e^{\nu b}\partial _{\mu }e_{\nu }{}^{a},  \label{eq33b}
\end{equation}
where $\omega _{\mu }{}^{ab}=-\omega _{\mu }{}^{ba}$ as consequence of the
metricity condition. This connection can be written as 
\begin{equation}
\omega _{\mu }{}^{ab}=\gamma _{\mu }{}^{ab}-K_{\mu }{}^{ba},  \label{eq33c}
\end{equation}
where 
\begin{equation}
K_{\mu }{}^{ba}=-K_{\mu }{}^{ab}=e^{\alpha a}e{}^{\beta b}K_{\mu \alpha
\beta },  \label{eq33d}
\end{equation}
${}$ is the contorsion tensor; while 
\begin{equation}
\gamma _{\mu }{}^{ab}=-\gamma _{\mu }{}^{ba}=e_{\mu i}{}\left(
C^{abi}-C^{bia}-C^{iab}\right) ,  \label{eq33f}
\end{equation}
is the Riemannian part of the connection $\omega _{\mu }{}^{ab}$ obtained
from the Ricci rotation coefficients $C^{abi}$ 
\begin{equation}
C_{ab}{}^{i}=e^{\mu }{}_{a}\left( x\right) e^{\nu }{}_{b}\left( x\right)
\partial _{\left[ \mu \right. }e_{\left. \nu \right] }{}^{i}.  \label{eq33g}
\end{equation}

The covariant derivative of the DKP field is then given by 
\begin{equation}
\nabla _{\mu }\psi =D_{\mu }\psi =\left( \partial _{\mu }+\frac{1}{2}\omega
_{\mu ab}S^{ab}\right) \psi ,  \label{eq46}
\end{equation}
and 
\begin{equation}
\nabla _{\mu }\overline{\psi }=D_{\mu }\overline{\psi }=\partial _{\mu }%
\overline{\psi }-\frac{1}{2}\omega _{\mu ab}\overline{\psi }S^{ab},
\label{eq47}
\end{equation}
where 
\begin{equation}
S_{ab}=\left[ \beta _{a},\beta _{b}\right] =\beta _{a}\beta _{b}-\beta
_{b}\beta _{a}.  \label{eq47b}
\end{equation}

Finally we will use semicolons to indicate the Riemannian part of the
covariant derivative of an object $A^{\nu a}{}$; i.e. its covariant
derivative in a Riemann manifold 
\begin{equation}
A^{\nu a}{}_{;\mu }=\partial _{\mu }A^{\nu a}+\gamma _{\mu
}{}^{a}{}_{b}A^{\nu b}+%
{\nu  \atopwithdelims\{\} \mu \alpha }%
A^{\alpha a}.  \label{eq62}
\end{equation}
Specifically for DKP field we have 
\begin{equation}
\psi _{;\mu }=\left( \partial _{\mu }+\frac{1}{2}\gamma _{\mu
ab}S^{ab}\right) \psi .  \label{eq62b}
\end{equation}

\subsection{Minimal coupling of DKP field}

Now we can consider the Lagrangian density of DKP field minimally coupled to
the ${\cal U}^{4}$ manifold 
\begin{equation}
{\cal L}=\sqrt{-g}\left[ \frac{i}{2}\left( \overline{\psi }\beta ^{\mu
}\nabla _{\mu }\psi -\nabla _{\mu }\overline{\psi }\beta ^{\mu }\psi \right)
-m\overline{\psi }\psi \right] ,  \label{eq48}
\end{equation}
where 
\begin{equation}
\beta ^{\mu }=e^{\mu }{}_{a}\beta ^{a},  \label{eq34}
\end{equation}
and satisfy 
\begin{equation}
\beta ^{\mu }\beta ^{\nu }\beta ^{\alpha }+\beta ^{\alpha }\beta ^{\nu
}\beta ^{\mu }=\beta ^{\mu }g^{\nu \alpha }+\beta ^{\alpha }g^{\nu \mu }.
\label{eq35}
\end{equation}
From the Lagrangian above we get the generalized DKP equation of motion 
\begin{equation}
i\beta ^{\mu }\nabla _{\mu }\psi +\frac{i}{2}K_{\mu b}{}^{\mu }\beta
^{b}\psi -m\psi =0,  \label{eq54}
\end{equation}
where we made use of 
\[
\partial _{\nu }\left( \sqrt{-g}e^{\nu }{}_{a}\right) =\sqrt{-g}\gamma
_{b}{}^{b}{}_{a}=\sqrt{-g}\gamma _{\mu }{}^{\mu }{}_{a},
\]
valid for a general Riemann-Cartan manifold\footnote{%
At this point we depart from the case presented in reference \cite{Lunardi 1}
since there we have $\partial _{\nu }\left( \sqrt{-g}e^{\nu }{}_{a}\right) =%
\sqrt{-g}\omega _{b}{}^{b}{}_{a}=\sqrt{-g}\omega _{\mu }{}^{\mu }{}_{a}$
because the manifold is riemannian.}. Some considerations about this result
are necessary. First of all, we can see from equation (\ref{eq54}) that the
minimal coupling procedure performed in the Lagrangian density leads to an
equation of motion which is not minimally coupled. This result is similar to
the one obtained in the case of Dirac's field, as can be seen in reference 
\cite{Sabbata}.

Moreover, the generalized DKP equation (\ref{eq54}) can be cast in the form%
\footnote{%
Here we have $\beta ^{\left[ a\right. }\beta ^{b}\beta ^{\left. c\right] }=%
\frac{1}{6}\left( \beta ^{a}\beta ^{b}\beta ^{c}+\beta ^{b}\beta ^{c}\beta
^{a}+\beta ^{c}\beta ^{a}\beta ^{b}-\beta ^{b}\beta ^{a}\beta ^{c}-\beta
^{a}\beta ^{c}\beta ^{b}-\beta ^{c}\beta ^{b}\beta ^{a}\right) $, the total
antisymmetrization of the indexes.} 
\begin{equation}
i\beta ^{a}\psi _{;a}+i\frac{3}{2}K_{abc}\beta ^{\left[ a\right. }\beta
^{b}\beta ^{\left. c\right] }\psi -\frac{i}{2}K_{abc}\beta ^{c}\beta
^{a}\beta ^{b}\psi -m\psi =0,  \label{eq54b}
\end{equation}
showing that, differently from the Dirac's field case \cite{Sabbata}, the
coupling to the contorsion tensor is not only to its totally antisymmetric
part due to the presence of the term in $K_{abc}\beta ^{c}\beta ^{a}\beta
^{b}$. Finally we must notice that there is an interaction with torsion. One
could think that this interaction would disappear when we select the spin 0
sector of the theory but, as we will see next, this does not happen.

\subsection{Comparison with the KG field}

Now we can use the projectors $P$ and $P^{a}$ that select the spin 0 sector
of the theory (see \cite{Umezawa,Lunardi 1,Lunardi 2}). From these
projectors defined in ${\cal M}^{4}$ we can construct the projectors in $%
{\cal U}^{4}$ as 
\begin{equation}
P^{\mu }=e^{\mu }{}_{a}P^{a}=e^{\mu }{}_{a}P\beta ^{a}=P\beta ^{\mu }.
\label{eq55}
\end{equation}
From the definitions above and the properties of $P$ and $P^{a}$ it is easy
to verify that 
\begin{equation}
P^{\mu }\beta ^{\nu }=Pg^{\mu \nu },\ PS^{\mu }{}^{\nu }=0,  \label{eq56}
\end{equation}
and it can also be seen that $P\nabla _{\mu }\psi =\nabla _{\mu }\left(
P\psi \right) $ and $P^{\nu }\nabla _{\mu }\psi =\nabla _{\mu }\left( P^{\nu
}\psi \right) $, where $\nabla _{\mu }\left( P\psi \right) =\partial _{\mu
}\left( P\psi \right) $ and $\nabla _{\mu }\left( P^{\nu }\psi \right)
=\partial _{\mu }\left( P^{\nu }\psi \right) +\Gamma _{\mu \beta }{}^{\nu
}P^{\beta }\psi $, i.e. $P\psi $ is derivated as a scalar and $P^{\nu }\psi $
as a 4-vector.

So, applying the operators $P^{\alpha }$\ and $P$ to the generalized DKP
equation (\ref{eq54}) we get, respectively, 
\begin{equation}
P^{\alpha }\psi =\frac{i}{m}\nabla ^{\alpha }\left( P\psi \right) +\frac{i}{%
2m}K_{\mu \nu }{}^{\mu }g^{\alpha \nu }\left( P\psi \right) ,  \label{eq57}
\end{equation}
and 
\begin{equation}
P\psi =\frac{i}{m}\nabla _{\alpha }\left( P^{\alpha }\psi \right) +\frac{i}{%
2m}K_{\mu \alpha }{}^{\mu }\left( P^{\alpha }\psi \right) .  \label{eq58}
\end{equation}

We must remember that $P\psi $ is a scalar and $P^{\alpha }\psi $ is a vector%
\footnote{%
Being rigorous, each component of the columns vectors $P\psi $ and $%
P^{\alpha }\psi $ is, respectively, a scalar and a vector.} so that we can
calculate their covariant derivatives $\nabla ^{\alpha }$ by considering
only the connection (\ref{eq32b}) acting on the Riemannian indexes \cite
{Lunardi 1}; specifically $\nabla _{\alpha }\left( P\psi \right) =\partial
_{\alpha }\left( P\psi \right) $ and $\nabla _{\alpha }\left( P^{\nu }\psi
\right) =\partial _{\alpha }\left( P^{\nu }\psi \right) +\Gamma _{\alpha \mu
}{}^{\nu }\left( P^{\mu }\psi \right) $. Combining equations (\ref{eq57})
and (\ref{eq58}) we get the equation of motion for the scalar field $P\psi $
as 
\begin{eqnarray}
\nabla _{\alpha }\nabla ^{\alpha }\left( P\psi \right)  &+&m^{2}\left( P\psi
\right) +\frac{1}{2}\nabla _{\alpha }\left( K_{\mu \nu }{}^{\mu }g^{\alpha
\nu }\left( P\psi \right) \right)   \nonumber \\
&+&\frac{1}{2}K_{\mu \alpha }{}^{\mu }\nabla ^{\alpha }\left( P\psi \right) +%
\frac{1}{4}K_{\mu \alpha }{}^{\mu }K_{\beta }{}^{\alpha \beta }\left( P\psi
\right) =0,  \label{eq59}
\end{eqnarray}
or in a more compact form 
\begin{equation}
\left( \nabla _{\alpha }+\frac{1}{2}K_{\mu \alpha }{}^{\mu }\right) \left(
\nabla _{\beta }+\frac{1}{2}K_{\mu \beta }{}^{\mu }\right) g^{\alpha \beta
}\left( P\psi \right) +m^{2}\left( P\psi \right) =0.  \label{eq59b}
\end{equation}

As we just mentioned above, the interaction with torsion does not disappear,
even after we selected the spin 0 sector of the DKP field. This interaction
is present both in the connection $\Gamma _{\mu \nu }{}^{\alpha }$ used in
the calculation of the covariant derivative $\nabla ^{\alpha }$ and in the
explicit presence of terms containing the contorsion tensor $K_{\mu \nu
}{}^{\alpha }$ in the equation above. This is in contrast with the result
obtained by performing the minimal coupling in the KG field Lagrangian
density, i.e. 
\begin{equation}
{\cal L}=\sqrt{-g}\left[ \partial ^{\mu }\varphi ^{\ast }\partial _{\mu
}\varphi -m\varphi ^{\ast }\varphi \right] ,  \label{eq60}
\end{equation}
that results in the equation of motion 
\begin{equation}
\left( \partial ^{\mu }\varphi \right) _{;\mu }+m^{2}\varphi =0,
\label{eq61}
\end{equation}
so that there is no interaction with torsion. Then we find that, in
Einstein-Cartan theory with minimal coupling procedure, DKP and KG fields
yield qualitatively different results: the first has an interaction with
torsion in its spin 0 sector while the second does not. So, differently from
others situations \cite{Lunardi 1,Lunardi 2,Fainberg 1,Fainberg 2,Fainberg 3}%
, DKP and KG theories for spin 0 particles are inequivalent in this context.
The origin of this inequivalence can be identified with the fact that the
free DKP field Lagrangian reduces, by a suitable choice of a rank 5
representation of the $\beta $ matrices, to the second order scalar field
Lagrangian \cite{Lunardi 2} 
\begin{equation}
{\cal L}=-\frac{1}{2}\left( \varphi ^{\ast }\square \varphi +\varphi \square
\varphi ^{\ast }\right) -m^{2}\varphi ^{\ast }\varphi ,  \label{eq61b}
\end{equation}
($\square =\partial ^{\mu }\partial _{\mu }$) and not the first order KG
Lagrangian. This is irrelevant in the free field case since these two
Lagrangians differ only by a complete divergence, what is also true in most
of the interaction cases. In the case of electromagnetic interaction, for
example, the minimally coupled DKP Lagrangian can be reduced to a second
order scalar field Lagrangian which is exactly the one obtained by
performing the minimal electromagnetic coupling in the second order free
Lagrangian (\ref{eq61b}) above. But this second order Lagrangian differs
from the minimally coupled first order KG Lagrangian only by a complete
divergence, so that there is a complete equivalence between them \cite
{Lunardi 2}.

As DKP Lagrangian with interaction can always be reduced to a second order
interacting scalar field Lagrangian, the equivalence between interacting DKP
Lagrangian and {\bf first order} KG Lagrangian depends on the equivalence
between this second order Lagrangian, obtained from the DKP one, and the
first order KG Lagrangian, i.e. whether they differ only by a complete
divergence. In General Relativity, as another example, there is such
equivalence \cite{Lunardi 1} because DKP Lagrangian can be reduced to a
second order scalar field Lagrangian, the same obtained by performing
minimal coupling in the Lagrangian (\ref{eq61b}), which differs from the
first order KG Lagrangian only by a complete divergence.

Similarly, in the case of Einstein-Cartan theory interaction, the specific
choice of $\beta $ matrices given in \cite{Lunardi 2} will give as result 
\begin{equation}
P\psi =\left( 
\begin{array}{c}
0_{4\times 1} \\ 
\psi _{4}
\end{array}
\right) ,\ P^{\mu }\psi =e^{\mu a}P_{a}\psi =e^{\mu a}{}\left( 
\begin{array}{c}
0_{4\times 1} \\ 
\psi _{a}
\end{array}
\right) ,  \label{eq61c}
\end{equation}
such that we will have, with the use of (\ref{eq57}), that the DKP field can
be written as 
\begin{equation}
\psi =\left( 
\begin{array}{c}
\frac{i}{\sqrt{m}}\left( \nabla _{a}\varphi +\frac{1}{2}K_{\mu a}{}^{\mu
}\varphi \right)  \\ 
\sqrt{m}\varphi 
\end{array}
\right) =\left( 
\begin{array}{c}
\frac{i}{\sqrt{m}}\left( \partial _{a}\varphi +\frac{1}{2}K_{\mu a}{}^{\mu
}\varphi \right)  \\ 
\sqrt{m}\varphi 
\end{array}
\right) ,  \label{eq61d}
\end{equation}
where $\varphi $ is a scalar field. If we compare this result with the
similar expression obtained in the case of electromagnetic interaction
(equation (54) of reference \cite{Lunardi 2}), we see that the expression
above for $\psi $ makes explicitly clear the existence of an interaction
between the scalar field and torsion. Moreover, one can expect a nonminimal
coupling since the derivative present in first four components of $\psi $ is
not the covariant derivative of Einstein-Cartan theory. Indeed, it can be
seen that the corresponding second order scalar field Lagrangian will be 
\begin{equation}
{\cal L}=\sqrt{-g}\left[ -\frac{1}{2}\left( \varphi ^{\ast }\nabla _{\mu
}\partial ^{\mu }\varphi +\varphi \nabla _{\mu }\partial ^{\mu }\varphi
^{\ast }\right) +\frac{1}{4}K_{\mu \alpha }{}^{\mu }K_{\beta }{}^{\alpha
\beta }\varphi ^{\ast }\varphi -m^{2}\varphi ^{\ast }\varphi \right] ,
\label{eq61e}
\end{equation}
which has equation (\ref{eq59}) as equation of motion but is not obtained by
minimal coupling from the free field Lagrangian (\ref{eq61b}) due to the
presence of the term $\frac{1}{4}K_{\mu \alpha }{}^{\mu }K_{\beta
}{}^{\alpha \beta }\varphi ^{\ast }\varphi $.

Now, if one performs an integration by parts in the minimally coupled first
order KG Lagrangian (\ref{eq60}) it is not possible to obtain a term
containing torsion since the derivative $\partial _{\mu }\sqrt{-g}$ will
only generate terms containing the Christoffel symbol (\ref{eq32c}). So, the
second order Lagrangian obtained in this way will not have interaction with
torsion, differently from the Lagrangian (\ref{eq61e}), where there is an
explicit interaction with torsion.

Summing up, the inequivalence between DKP and KG fields in Einstein-Cartan
theory is due to the fact that the minimal coupling in DKP Lagrangian
corresponds to a nonminimal coupling\footnote{%
The fact that the coupling is nonminimal is not fundamental. Even if the
second order scalar Lagrangian obtained from DKP one were simply that
obtained by performing minimal coupling in Lagrangian (\ref{eq61b}) there
would be an interaction with torsion and a inequivalence with the first
order KG Lagrangian.} in a second order scalar Lagrangian, what results in
an interaction with torsion, contrary to the minimal coupling in the first
order KG Lagrangian. We should also remember the necessity of using the
symmetric DKP Lagrangian (\ref{eq48}) and not the nonsymmetric one, i.e. 
\begin{equation}
{\cal L}=\sqrt{-g}\left[ i\overline{\psi }\beta ^{\mu }\nabla _{\mu }\psi -m%
\overline{\psi }\psi \right] ,  \label{eq64b}
\end{equation}
because they are not equivalent, contrary to what happens in the cases of
General Relativity or electromagnetic interactions. This is due to the fact
that the symmetric Lagrangian can not be obtained from the nonsymmetric one
by an integration by parts because the derivative $\partial _{\mu }\sqrt{-g}$
will not generate the contorsion terms necessary to construct the complete
Riemann-Cartan connection, present in the covariant derivative of $\overline{%
\psi }$ contained in the symmetric Lagrangian. Moreover, Lagrangian (\ref
{eq64b}) would result in a equation of motion for $\overline{\psi }$ field
which can not be obtained by conjugation from the equation for $\psi $. But
these results are not particularities of DKP field but a general
characteristic of fields in Riemann-Cartan manifolds, as can be seen through
a similar analysis, for example, of the second order KG or Dirac
Lagrangians. We also should mention that there would be no inequivalence if
the minimal coupling where performed in the equation of motion instead of in
the Lagrangian. In this case the generalized DKP equation in Riemann-Cartan
space is given by 
\begin{equation}
i\beta ^{\mu }\nabla _{\mu }\psi -m\psi =0,  \label{eq63}
\end{equation}
which, by means of the projectors $P$ and $P^{\alpha }$, provides 
\begin{equation}
\nabla ^{\mu }\nabla _{\mu }\left( P\psi \right) -m\left( P\psi \right)
=\nabla ^{\mu }\partial _{\mu }\left( P\psi \right) -m\left( P\psi \right)
=0.  \label{eq64}
\end{equation}
Since $P\psi $ is a scalar, this result is completely equivalent to the
minimal coupling performed in the KG equation of motion, both cases
resulting in a coupling with torsion via the contorsion tensor present in
the covariant derivative $\nabla ^{\mu }$. As can be seen from these
results, performing minimal coupling in the equation of motion gives
different results from the same coupling performed in the Lagrangian, for
both DKP and KG fields; what is, once more, a characteristic of fields in
Riemann-Cartan manifolds: the same occurs with Dirac field \cite{Sabbata}.

Finally, it is necessary to mention that this result is qualitatively
different from the torsion coupling with the KG field described in reference 
\cite{Andrade 1} in the context of Teleparallel Description of Gravity. In
the Teleparallel formalism the General Relativity's description of gravity
as a curvature effect is replaced by a description in terms of a flat space
with torsion (Weitzenb\"{o}ck spacetime) so that the interaction with
torsion is necessary to exist a gravitational interaction at all. The result
we have just obtained above, on the contrary, describes the interaction of
the spin 0 sector of DKP field with torsion in the context of
Einstein-Cartan theory, where both torsion and curvature are present. This
point will be made more clear in the next section, where we will analyse the
interaction of the DKP field with torsion in the context of Teleparallel
Description of Gravity.

\section{DKP field in Teleparallel Gravity}

Here we will analyse the DKP field in the context of Teleparallel version of
General Relativity \cite{Hayashi}. In this case we have a Weitzenb\"{o}ck
space-time; a particular case of Riemann-Cartan space-time constrained to
have zero curvature or, equivalently, to satisfy the teleparallel condition
of the tetrad field 
\begin{equation}
\stackrel{c}{\nabla }_{\mu }e_{\nu }{}^{a}=\partial _{\mu }e_{\nu
}{}^{a}-\Gamma _{\mu \nu }{}^{\alpha }e_{\alpha }{}^{a}=0.  \label{eq65}
\end{equation}
In the expression above $\stackrel{c}{\nabla }_{\mu }$ is the covariant
derivative with respect only to the Riemann-Cartan connection and differs
from the total covariant derivative $\nabla _{\mu }$, given in equation (\ref
{eq31}), by the absence of the $\omega _{\mu }{}^{ab}$ spin connection. This
condition implies that the spin connection $\omega _{\mu }{}^{ab}$ vanishes
identically and that the connection $\Gamma $ is the (nonsymmetric) Cartan
connection 
\begin{equation}
\Gamma _{\mu \nu }{}^{\alpha }=e^{\alpha }{}_{i}\partial _{\mu }e_{\nu
}{}^{i}.  \label{eq66}
\end{equation}
We also have 
\begin{equation}
\partial _{\mu }e=ee^{\nu }{}_{a}\partial _{\mu }e_{\nu }{}^{a}=e\Gamma
_{\mu \nu }{}^{\nu },  \label{eq66b}
\end{equation}
where $e=\det e_{\nu }{}^{a}$, and combining equations (\ref{eq65}) and (\ref
{eq32b}) we find that 
\begin{equation}
\stackrel{r}{\nabla }_{\mu }e_{\nu }{}^{a}=\partial _{\mu }e_{\nu }{}^{a}-%
{\alpha  \atopwithdelims\{\} \mu \nu }%
e_{\alpha }{}^{a}=-K_{\mu \nu }{}^{\alpha }e_{\alpha }{}^{a},  \label{eq67}
\end{equation}
\ where $\stackrel{r}{\nabla }_{\mu }$ indicates the covariant derivative
with respect to the Christoffel connection, i.e. the connection of General
Relativity. From the torsionless case of equation (\ref{eq33b}) we can also
write the spin connection $\stackrel{r}{\omega }_{\mu }{}^{ab}$ of General
Relativity as 
\begin{equation}
\stackrel{r}{\omega }_{\mu }{}^{ab}=e^{\nu b}%
{\alpha  \atopwithdelims\{\} \mu \nu }%
e_{\alpha }{}^{a}-e^{\nu b}\partial _{\mu }e_{\nu }{}^{a}=-e^{\nu b}%
\stackrel{r}{\nabla }_{\mu }e_{\nu }{}^{a}=e^{\nu b}K_{\mu \nu }{}^{\alpha
}e_{\alpha }{}^{a}=-K_{\mu }{}^{ab}.  \label{eq68}
\end{equation}

Now we must notice that the use of covariant derivative (\ref{eq46}), with $%
\omega _{\mu }{}^{ab}=0$ as consequence of teleparallel condition, will not
provide an equivalence with General Relativity, as can also be seen in the
case of Dirac field addressed in reference \cite{Andrade}. But, following
the procedure described in this reference, the covariant derivative $%
\stackrel{r}{\nabla }_{\mu }\psi =$ $\stackrel{r}{D}_{\mu }\psi $ of the DKP
field in General Relativity, i.e. in a Riemann manifold, can be written in
terms of teleparallel quantities. To do this, first we must notice that such
derivative can be obtained from the torsionless case of equation (\ref{eq46}%
) and is given by 
\begin{equation}
\stackrel{r}{\nabla }_{\mu }\psi =\stackrel{r}{D}_{\mu }\psi =\left(
\partial _{\mu }+\frac{1}{2}\stackrel{r}{\omega }_{\mu ab}S^{ab}\right) \psi
.  \label{eq69}
\end{equation}

So, with the help of equation (\ref{eq68}) above, it can be rewritten in
terms of teleparallel quantities as 
\begin{equation}
\stackrel{r}{\nabla }_{\mu }\psi =\stackrel{r}{D}_{\mu }\psi =\left(
\partial _{\mu }-\frac{1}{2}K_{\mu \alpha \beta }S^{\alpha \beta }\right)
\psi .  \label{eq70}
\end{equation}
Analogously, for $\overline{\psi }$ we have 
\begin{equation}
\stackrel{r}{\nabla }_{\mu }\overline{\psi }=\stackrel{r}{D}_{\mu }\overline{%
\psi }=\partial _{\mu }\overline{\psi }+\frac{1}{2}K_{\mu \alpha \beta }%
\overline{\psi }S^{\alpha \beta },  \label{eq71}
\end{equation}
so that the minimally coupled Lagrangian is written in terms of teleparallel
structure as 
\begin{equation}
{\cal L}=e\left[ \frac{i}{2}\left( \overline{\psi }\beta ^{\mu }\stackrel{r}{%
\nabla }_{\mu }\psi -\stackrel{r}{\nabla }_{\mu }\overline{\psi }\beta ^{\mu
}\psi \right) -m\overline{\psi }\psi \right] ,  \label{eq72}
\end{equation}
or 
\begin{equation}
{\cal L}=e\left[ \frac{i}{2}\left( \overline{\psi }\beta ^{\mu }\left(
\partial _{\mu }\psi -\frac{1}{2}K_{\mu \alpha \beta }S^{\alpha \beta }\psi
\right) -\left( \partial _{\mu }\overline{\psi }+\frac{1}{2}K_{\mu \alpha
\beta }\overline{\psi }S^{\alpha \beta }\right) \beta ^{\mu }\psi \right) -m%
\overline{\psi }\psi \right] ,  \label{eq73}
\end{equation}
from which we obtain the equation of motion 
\begin{equation}
i\beta ^{\mu }\stackrel{r}{\nabla }_{\mu }\psi -m\psi =0.  \label{eq76}
\end{equation}
Now the resulting coupled equation of motion is exactly the one obtained by
performing the minimal coupling in the equation of motion. The above
equation of motion is exactly the one obtained in General Relativity \cite
{Lunardi 1}, but here is written in terms of teleparallel quantities as we
can see from equation (\ref{eq70}). Moreover, an identical result is
obtained by performing the minimal coupling procedure directly in the
equation of motion. But this identity is expected since it occurs in General
Relativity and we are dealing with its teleparallel equivalent.

\subsection{Comparison with the KG field}

Due to properties of teleparallel structure, the operators $P$ and $P^{\nu }$
also have the properties $P\stackrel{r}{\nabla }_{\mu }\psi =\stackrel{r}{%
\nabla }_{\mu }\left( P\psi \right) $ and $P^{\nu }\stackrel{r}{\nabla }%
_{\mu }\psi =\stackrel{r}{\nabla }_{\mu }\left( P^{\nu }\psi \right) $, with 
$\stackrel{r}{\nabla }_{\mu }\left( P\psi \right) =\partial _{\mu }\left(
P\psi \right) $ and $\stackrel{r}{\nabla }_{\mu }\left( P^{\nu }\psi \right)
=\partial _{\mu }\left( P^{\nu }\psi \right) +\stackrel{r}{\Gamma }_{\mu
\beta }{}^{\nu }P^{\beta }\psi $, as in General Relativity. Consequently,
applying the operators $P$ and $P^{\nu }$ to the equation of motion (\ref
{eq76}) results in 
\begin{equation}
P\psi =\frac{i}{m}\stackrel{r}{\nabla }_{\mu }\left( P^{\mu }\psi \right) ,
\label{eq77}
\end{equation}
and 
\begin{equation}
P^{\lambda }\psi =\frac{i}{m}\stackrel{r}{\nabla }^{\lambda }\left( P\psi
\right) =\frac{i}{m}\partial ^{\lambda }\left( P\psi \right) ,  \label{eq78}
\end{equation}
respectively. Combining these results we get 
\begin{equation}
\stackrel{r}{\nabla }_{\mu }\stackrel{r}{\nabla }^{\mu }\left( P\psi \right)
+m^{2}P\psi =\stackrel{r}{\nabla }_{\mu }\partial ^{\mu }\left( P\psi
\right) +m^{2}P\psi =0.  \label{eq79}
\end{equation}
This result is exactly what is obtained in the teleparallel description of
the KG field presented in reference \cite{Andrade 1}, as it can be seen by
writing this last equation, with the help of equation (\ref{eq32b}), as 
\begin{equation}
\partial _{\mu }\partial ^{\mu }\left( P\psi \right) +\left( \Gamma _{\mu
\alpha }{}^{\mu }+K_{\mu \alpha }{}^{\mu }\right) \partial ^{\alpha }\left(
P\psi \right) +m^{2}P\psi =0,  \label{eq80}
\end{equation}
\begin{equation}
{\cal D}_{\mu }\partial ^{\mu }\left( P\psi \right) +m^{2}P\psi =0,
\label{eq81}
\end{equation}
where ${\cal D}_{\mu }=\partial _{\mu }+\Gamma _{\mu }+K_{\mu }=\nabla _{\mu
}+K_{\mu }$, being $\Gamma $ the Cartan connection, given by equation (\ref
{eq66}), and $K$ the corresponding contorsion tensor, as given by equation (%
\ref{eq32d}). The derivative ${\cal D}_{\mu }$ is the teleparallel version
of General Relativity's covariant derivative given by equation (26) of
reference \cite{Andrade 1}, so equation (\ref{eq81}) above is exactly the
same as equation (33) of this reference\footnote{%
We must remember that the definitions here used for the connections indexes
and for the contorsion tensor are different from those in the mentioned
reference.}. Furthermore we can also write equation (\ref{eq81}) as 
\begin{equation}
\nabla _{\mu }\partial ^{\mu }\left( P\psi \right) +m^{2}P\psi =-K_{\mu
\alpha }{}^{\mu }\partial ^{\alpha }\left( P\psi \right) ,  \label{eq82}
\end{equation}
which is also identical to equation (34) of reference \cite{Andrade 1}. So
we can see that, in the teleparallel description of General Relativity, the
spin 0 sector of the DKP field is exactly equivalent to KG field.

\section{Conclusions and comments}

We have analysed the minimal coupling of the DKP field with torsion in two
contexts. First, we have shown that in a Riemann-Cartan manifold there is an
interaction with torsion that persists even when we select the spin 0 sector
of the theory. So, in this context, the DKP field provides a coupling of
spin 0 particles with torsion when we perform minimal coupling; a result
different from that obtained when we describe the spin 0 field using the KG
field\footnote{%
We must stress that this will also occur with the spin 0 sector of any
redutible representation of $\beta $ matrices.}. This difference comes, as
we mentioned before, from the fact that, in a Riemann-Cartan manifold, the
DKP Lagrangian corresponds to a second order scalar field Lagrangian, given
by equation (\ref{eq61e}), which is not equivalent to the first order KG
Lagrangian given by equation (\ref{eq60}); a result that is different from
the usual equivalence obtained in the case of other interactions. It should
also be noticed that the DKP coupling here is purely classical and has no
relation with quantum effects; as those that produce the coupling of
quantized electromagnetic field with torsion via second order terms in
pertubative series \cite{Sabbata 2}.

Next we have shown that in the case of teleparallel description of General
Relativity, constructed in a Weitzenb\"{o}ck space-time, there is a perfect
agreement between the results of the DKP field and the KG field, even in the
presence of an interaction of the spin 0 sector with torsion, as it can be
seen in equations (\ref{eq79}), (\ref{eq81}) or (\ref{eq82}) above. This
represents no contradiction with the results of Riemann-Cartan case since
here we are dealing with a Weitzenb\"{o}ck manifold to construct a
teleparallel equivalent of General Relativity. Since such agreement occurs
in General Relativity it must also occur in its teleparallel equivalent. The
point here is that the origin of the inequivalence does not exist anymore,
since the teleparallel condition, equation (\ref{eq65}), implies a relation
between the contorsion tensor and the Christoffel symbol, since the
contributions of each to the curvature must cancel. Such relation is what
makes possible to write the Riemannian quantities of General Relativity in
terms of the quantities of Weitzenb\"{o}ck space-time and also makes the
first order KG Lagrangian in such description again equivalent to the second
order scalar Lagrangian obtained from DKP one, since now one can be obtained
from the other by an integration by parts. Consequently, the equivalence
between both theories will be restored.

We must also stress that the coupling with torsion has completely different
physical meanings in each of the cases analysed above. In the teleparalell
description we exchanged the description of General Relativity in terms of
curvature and metric for a description in terms of torsion and tetrads,
without curvature. In this case the interaction with torsion substitutes the
interaction with curvature, both in the case of the DKP field described here
or in the case of the KG field \cite{Andrade 1}. Diversely, in the
Riemann-Cartan manifold we have both curvature and torsion and the minimally
coupled KG field does not interact with torsion in this context while the
DKP field does. Finally, we would like to emphasize that the {\it projected
equations} (\ref{eq59}), (\ref{eq59b}), (\ref{eq64}) and (\ref{eq82}), which
shows explicitly the scalar sector of the DKP field, are used only to
provide a comparison with the KG field. It is the linearity of the DKP
equation that, despite the complicate algebra of $\beta $ matrices, can be
an advantage when studying spin 0 fields in curved space-times and for this
it is the DKP equation in the forms obtained here that should be used.
Moreover, the natural coupling of the DKP field to torsion in
Einstein-Cartan theory makes it an interesting alternative to study
interaction of spin 0 fields with torsion without introducing nonminimal
couplings.

\section{Acknowledgments}

J.T.L. and B.M.P. would like to thank CAPES's PICDT program and CNPq,
respectively, for partial support. R.G.T. thanks FAPESP (grant 00/02263-9)
for full support. The authors thank Prof. J. G. Pereira for his comments on
the Teleparallel Gravity.

\end{document}